\documentclass[lettersize,journal]{IEEEtran}
\usepackage{amsmath,amsfonts}
\usepackage{algorithmic}
\usepackage{algorithm}
\usepackage{array}
\usepackage[caption=false,font=normalsize,labelfont=sf,textfont=sf]{subfig}
\usepackage{textcomp}
\usepackage{stfloats}
\usepackage{url}
\usepackage{verbatim}
\usepackage{graphicx}
\usepackage{cite}
\usepackage{balance}
\usepackage{multirow}
\usepackage{siunitx}

\begin{document}

\title{A Sample Article Using IEEEtran.cls\\ for IEEE Journals and Transactions}
\title{\LARGE Sparse-on-Dense: Area and Energy-Efficient Computing of Sparse Neural Networks on Dense Matrix Multiplication Accelerators}

\author{Hyunsung Yoon$^*$,~\IEEEmembership{Graduate Student Member,~IEEE,}
        Sungju Ryu$^*$,~\IEEEmembership{Member,~IEEE,}
        Jae-Joon Kim,~\IEEEmembership{Member,~IEEE}% <-this % stops a space
%\thanks{This paper was produced by the IEEE Publication Technology Group. They are in Piscataway, NJ.}% <-this % stops a space
%\thanks{Manuscript received April 19, 2021; revised August 16, 2021.}}
        \thanks{Hyunsung Yoon is with the Department of Convergence IT Engineering, Pohang University of Science and Technology, Pohang, Republic of Korea.}
        \thanks{(email: hyunsung.yoon@postech.ac.kr).}
        \thanks{Sungju Ryu is with the Department of System Semiconductor Engineering, Sogang University, Seoul, Republic of Korea.}
        \thanks{(email: sungju@sogang.ac.kr).}
        \thanks{Jae-Joon Kim is with the Department of Electrical and Computer Enginnering, Seoul National University, Seoul, Republic of Korea.}
        \thanks{(email: kimjaejoon@snu.ac.kr, Corresponding Author).}
        % \thanks{Manuscript received XXX XX, 2024; revised YYY YY, 2024.}
}
% The paper headers
% \markboth{IEEE Transactions on Computer-Aided Design of Integrated Circuits \& Systems,~Vol.~XX, No.~X, YY~2024}
% {Yoon \MakeLowercase{\textit{et al.}}: Sparse-on-Dense: Area and Energy-Efficient Computing of Sparse NNs on Dense MM Accelerators}

%\IEEEpubid{0000--0000/00\$00.00~\copyright~2021 IEEE}
% Remember, if you use this you must call \IEEEpubidadjcol in the second
% column for its text to clear the IEEEpubid mark.

\maketitle
\def\thefootnote{*}\footnotetext{These authors contributed equally to this work.}

\begin{abstract}
As the size of Deep Neural Networks (DNNs) increases dramatically to achieve high accuracy, the DNNs require a large amount of computations and memory footprint.
Pruning, which produces a sparse neural network, is one of the solutions to reduce the computational complexity of neural network processing.
To maximize the performance of the computations with such compressed data, dedicated sparse neural network accelerators have been introduced, but complex circuits for matching the indices of non-zero inputs/weights cause large overhead in area and power of processing elements (PEs). 
The sparse PE becomes significantly larger than the dense PE, which raises an interesting question for designers; \textbf{“Given the area, isn’t it better to use larger number of dense PEs despite the low utilization in sparse matrix computations?”}  
In this paper, we show that the answer is “yes”, and demonstrate an area and energy-efficient method for sparse neural network computing on dense-matrix multiplication hardware accelerators (Sparse-on-Dense).
\end{abstract}


\section{Introduction}
\IEEEPARstart{D}{eep} Neural Networks (DNNs) are becoming more common in everyday life, thanks to their superior performance in various applications such as computer vision \cite{krizhevsky2012alexnet, simonyan2014vgg}, natural language processing \cite{brown2020language_gpt3, devlin2018bert}, and personalization \cite{naumov2019dlrm}.
The size of DNN models is growing dramatically to expand the range of DNN applications and improve accuracy for those services \cite{brown2020language_gpt3}.
As the number of parameters in each DNN model increases, the memory footprint, number of memory accesses, and computational complexity also increase.

Many data compression techniques, such as quantization and pruning \cite{han2015deep}, have been proposed to alleviate such a model complexity \cite{cheng2017model}. Quantization reduces the amount of data by lowering the bit precision while keeping the number of weight and input elements the same. 
 Pruning, in contrast, reduces the number of the parameters by approximating near-zero values to zero. 
 The number of multiply-accumulate (MAC) operations is significantly reduced by skipping the \textit{multiply-by-zero} computations. 

 \begin{figure}[t]
  \centering
  \includegraphics[width=\linewidth]{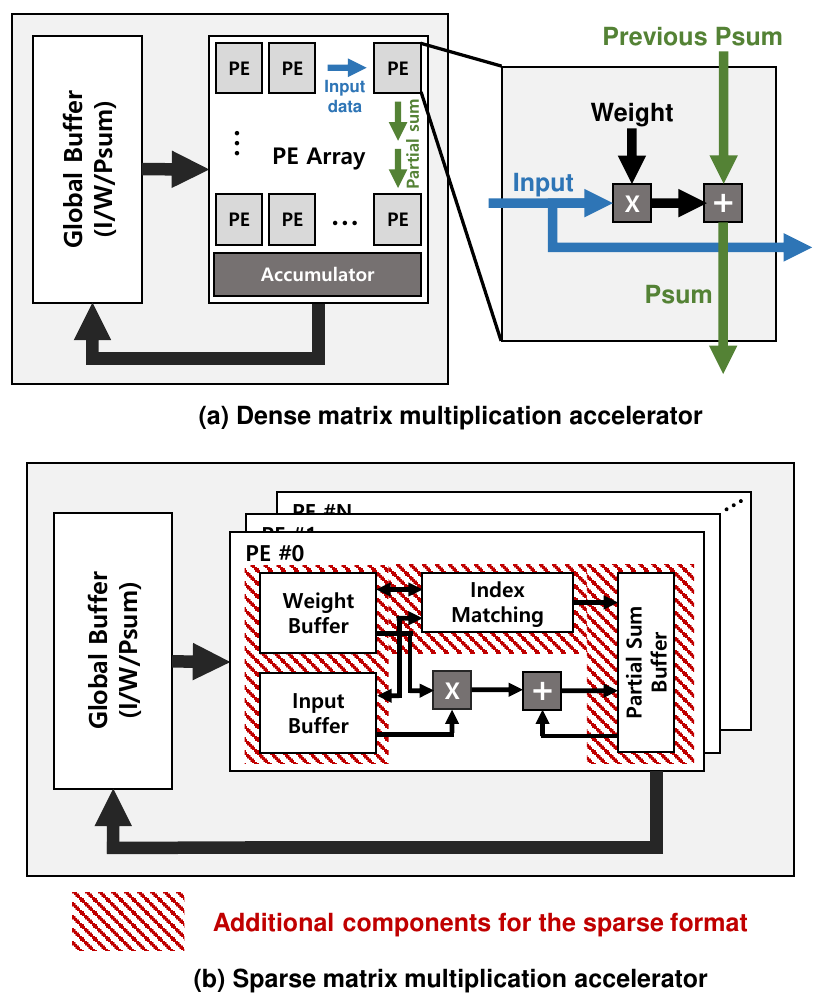}
  \caption{Architectures of the neural network accelerators. (a) Dense matrix multiplication accelerator and (b) sparse matrix multiplication accelerator.}
%   \Description{Architecture for the neural network acceleration differs by the density}
  \label{prev_arch}
\end{figure}

Data compression formats are commonly used to represent sparse matrices in order to take advantage of the benefits of sparse neural networks. The compressed sparse column (CSC) format, for example, is made up of three parts: the value of the non-zero elements, an index that represents the row index of the value, and a pointer which indicates the number of nonzero elements in each column. By adopting such compression formats in sparse matrices, the model complexity of the sparse neural networks can be reduced. 

\begin{figure*}[t]
  \centering
  \includegraphics[width=\linewidth]{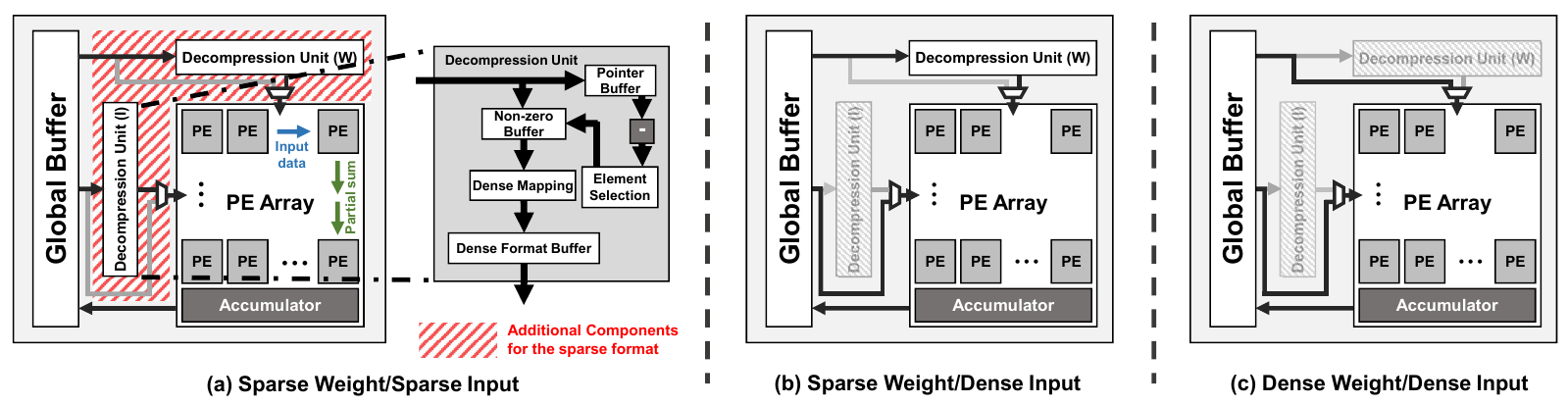}
  \caption{Overall architecture of Sparse-on-Dense and its operation mode depending on the density of the data. The dense data bypasses the decompression unit.}
  %\Description{Overall architecture to support Sparse-on-dense method. The PE array is adopted from the systolic array of Google's TPU \cite{isca_tpu}}.
  \label{sod_arch}
\end{figure*}

Many hardware accelerators tailored to sparse neural networks \cite{han2017ese, parashar2017scnn, zhang2020snap} have been introduced to maximize the advantages of the compressed data formats. However, the non-zero data indices typically have the irregular pattern. As a result, additional index-matching logic is required to match non-zero inputs/weights, and large buffers are used in processing elements (PEs) to balance the number of non-zero values between the PEs.

The area of the PE that handles the sparse networks (sparse-PE) becomes significantly large, especially compared to area of the PE that handles the dense networks (dense-PE). 
As a result, there is a great need to evaluate whether using large sparse-PE with higher utilization is better than using small dense-PE with lower utilization for computing sparse networks for a given area to the PE array.

To compare the design choices, we first need to develop a scheme to feed the sparse data into the dense PE array.
In this paper, we present the Sparse-on-Dense hardware architecture, which   takes advantage of both 1) simple PE structure of the dense matrix multiplication accelerators and 2) reduced access to external memory of the sparse matrix multiplication accelerators.
The evaluation results show that mapping the sparse network onto dense PE arrays produces the higher performance for a given area for PE array than using complex sparse PE array. Sparse-on-Dense architecture also allows users to freely run dense or sparse networks with unstructured pruning on the neural processing unit without worrying about hardware constraint.

\section{Preliminary}

% A lot of dedicated hardware accelerators have been proposed to process DNNs efficiently. The accelerators have different design points depending on the density of the neural network that each hardware targets. This section outlines the characteristics of the neural network accelerators by the density of the processing data.
\subsection{Dense Neural Network Accelerator}
%\subsection{Sparse Neural Networks on Dense Matrix Multiplication Accelerator}
%%figure showing systolic dataflow and simple pe architecture
% Dense neural network accelerators are usually introduced for the efficient matrix multiplication, which is a fundamental operation of the DNNs. The matrix multiplication is performed by repeating multiplication between matrix elements and accumulation of the multiplication results. During this repeated computation, each element is used several times, being multiplied by the different elements. Furthermore, the elements required for sequential multiplication can be accessed serially. Therefore, the accelerators for dense neural networks usually adopt special dataflow that can reduce the data movement and maximize data reuse.

Dense matrix multiplication accelerators are frequently adopted in efficient matrix multiplication which is a fundamental operation of the DNNs. 
% Figure \ref{prev_arch}a shows the general hardware architecture of the dense neural network accelerator implementing the systolic dataflow. The systolic dataflow is introduced by {\bfseries Google’s TPU}\cite{isca_tpu}, one of the popular dense neural network accelerators. The accelerator includes large global buffers for the input and weight data, an accumulator and a processing element (PE) array. Each PE consist of a multiplication and accumulation (MAC) unit and some buffers for the weight, input and a partial sum. The systolic dataflow of the PE array maximizes data reuse by shifting data in one PE to the neighbor PEs. While the weight data stays in the buffer, the input data moves to the right-side PE that requires the same input data. When the element multiplication in the PE ends, the result is accumulated with the partial sum coming from the upper-side PE and stored in the buffer. As the weight in the PE does not change during the computation, the weight matrix should be tiled in the same size with the PE array. By using the systolic dataflow, the hardware accelerator shows high computational throughput with a simple PE architecture.
Fig. \ref{prev_arch}a shows a popular weight-stationary systolic array architecture (e.g.: Google TPU \cite{isca_tpu}).
%{\bfseries Google’s TPU}\cite{isca_tpu}. 
The accelerator includes large global buffers for the input and weight data, an accumulator and a PE array. Each PE consists of a MAC unit and buffers for input, weight, and partial sum (psum). The systolic dataflow of the PE array maximizes data reuse by shifting inputs/psums in one PE to the neighbor PEs.

\begin{figure}[t]
  \centering
  \includegraphics[width=\linewidth]{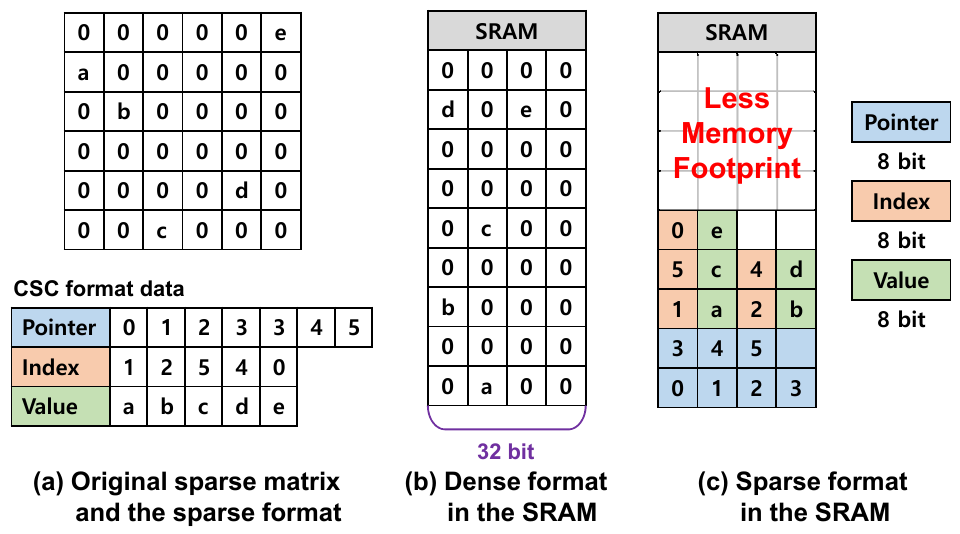}
  \caption{Different memory usage depending on the data format.}
  %\Description{The sparse format utilizes less memory.}
  \label{sparse_form_exp}
\end{figure}

% While TPU showed efficient matrix multiplication architecture for DNNs, the appearance of the sparse neural networks made the neural network accelerators enter another phase. Because the systolic PE array is optimized for the dense applications, there is an architectural limitation in handling the sparse data. Therefore, the hardware designs specialized for the sparse matrix have been researched.

While the systolic array is a simple and efficient matrix multiplication architecture for DNNs, it does not exploit the low computational density and the small number of memory accesses of the sparse neural network computation.
As a result, specialized hardware for the sparse matrix multiplication have been studied.

%{\bfseries Google’s TPU} is one of the popular dense neural network accelerators. TPU includes large unified buffer for the input data and a processing element (PE) array. Each PE consist of a multiplication and accumulation (MAC) unit and some buffers The systolic dataflow of the PE array maximizes data reuse by shifting data in one PE to the neighbor PEs. The input data moves to right-side PE that requires the same input data. The partial sum in one PE goes to downside PE and makes new partial sum after being accumulated with the multiplication result of the downside PE. With this dataflow, TPU shows high computational throughput with a simple PE architecture.

% \subsection{Sparse Neural Network Accelerator}
%\subsection{Sparse Neural Networks on Hardware Specialized in Sparse Matrix Multiplication}
\subsection{Sparse Neural Network Accelerator}
\label{prelim_sparse}
%%figure showing complex pe architecture of sparse nn accelerators
% The reduction in the weights and the computation from the sparsity has brought out the hardware implementation of the sparse neural networks, which has some challenges that the sparsity provokes. One well-known problem is the load imbalance among PEs. As the distribution of the left non-zero elements in the original matrix is random, the number of the weights assigned to each PE can be different over the PEs. Therefore the utilization decreases with the stalls caused by the imbalanced workloads. Furthermore, memory accesses for the input data is not predictable. Because the weights in each PE have the different indices, the input elements required at the same time is arbitrary. Even more, in the case of the sparse convolutional neural networks (CNNs), the sparsity is added to the input elements because of the ReLU activation function.

\textcolor{blue}{
\begin{table}[t]
\centering
%\footnotesize
%\setlength\tabcolsep{6pt}
\caption{Sparsity in the sparse neural network accelerators.}
\begin{tabular}{ |c||c|c|c| }
\hline
  & Sparsity (Weight)  & Sparsity(Input) \\
\hline
ESE~\cite{han2017ese} & Unstructured & Dense \\
SNAP~\cite{zhang2020snap} & Unstructured & Unstructured \\
SCNN~\cite{parashar2017scnn} & Unstructured & Unstructured \\
SIGAM~\cite{qin2020sigma} & Unstructured & Unstructured \\
STA~\cite{liu2020sta} & Structured (4/8) & Dense \\
S2TA~\cite{liu2022s2ta} & Structured (4/8) & Structured (1-5/8) \\
Ours (Sparse-on-Dense) & Unstructured & Unstructured \\
\hline
\end{tabular}
\label{table_sparsehw}
\end{table}
}

Sparse neural networks efficiently achieve the memory saving and reduction in the number of MAC operations, but they have some challenges; irregular non-zero index pattern, load imbalance between the PEs. So, many previous works have tried to solve the limitations.

\begin{figure*}[t]
  \centering
  \includegraphics[width=\linewidth]{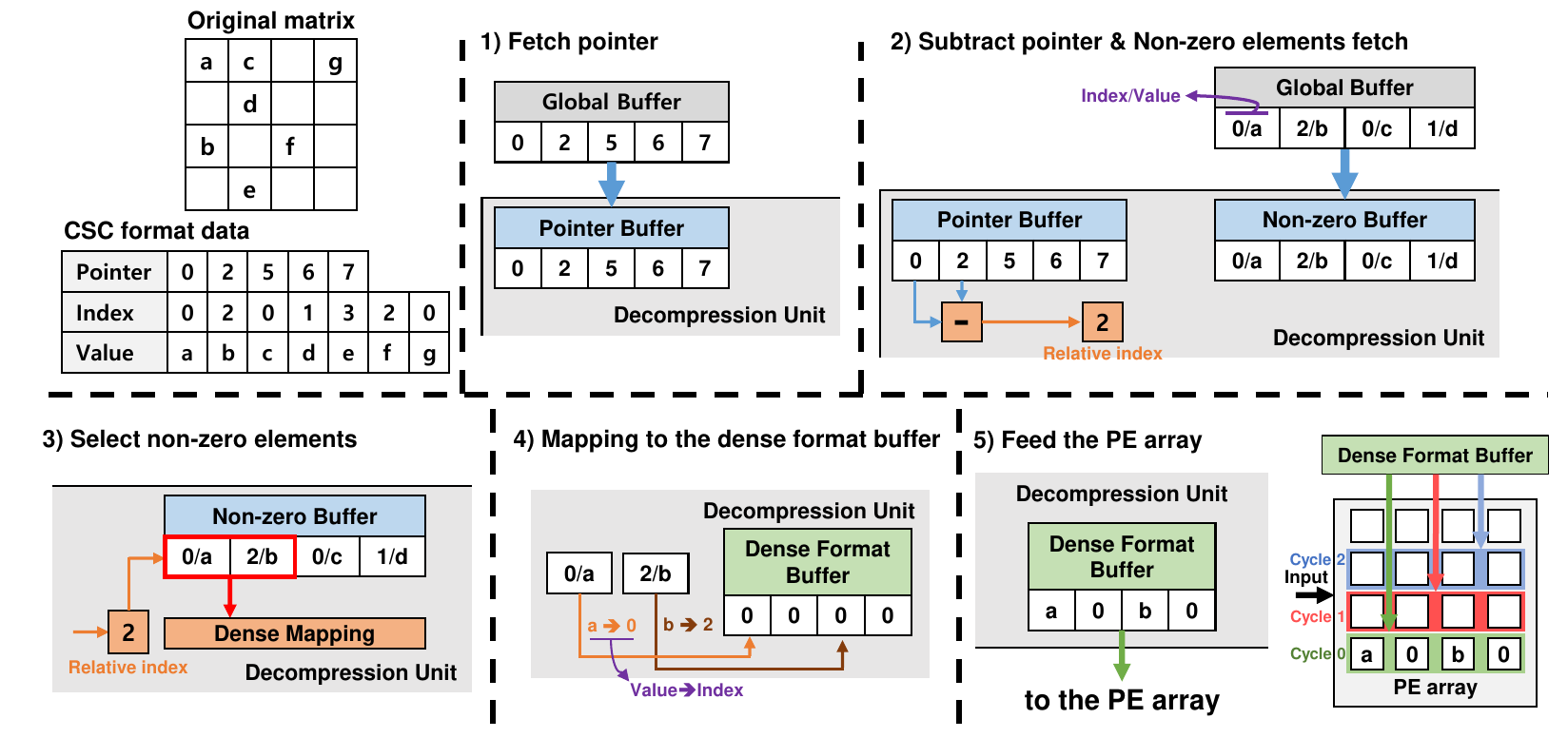}
   \vspace{-5mm}
  \caption{Processing steps to decompress the sparse weight encoded in the CSC format. Inputs are decompressed in the same manner as the weights.}
  \vspace{-5mm}
  %\Description{The way the decompression unit works for CSC.}
  \label{decomp_step}
\end{figure*}

% To address the challenges, the sparse neural network accelerators usually have different PE architecture from the dense neural network accelerator (Figure \ref{prev_arch}b). The major difference is an index matching unit. Because each PE requires different input elements, the sparse accelerators adopt index matching unit to find index matched pairs among the multiple weight data and the multiple input data. Therefore, the size of the weight buffer and the input buffer should be big enough to hold a number of data. AS the number of index matching pair is irregular, the partial sum buffer also has a large capacity to keep as many partial sums as possible.

%Beside the weight sparsity, made on purpose, CNN also has zero activations because of ReLU activation function. This activation sparsity gives additional randomness to the index matching for the matrix multiplication. Therefore, sparse CNN accelerators are generally designed to support dual sparsity using special unit to find a pair of the index-matched data.

{\bfseries ESE}~\cite{han2017ese} is a sparse neural network accelerator for LSTM neural network. Because the input data is a dense voice vector, ESE is specialized for the multiplication between the sparse weight and the dense input data. While the size of the weight buffer and the size of the partial sum buffer remain large, the dense input data allows each PE of ESE to have small input FIFO instead of the input buffer.
% When the input data from the FIFO has the same index with the sparse weight, the PE performs an element multiplication.
By using the FIFO, each weight index in PEs can be compared with multiple input indices in the FIFO, thereby the PE consumes the weights fast and mitigates the stalls.
When the input data from the FIFO has the same index as the index of the sparse weight, the PE multiplies the input by the matched non-zero weight.
% Considering the workload imbalance over the PEs, the input FIFO in each PE has different inputs to mitigate the stalls.

{\bfseries SNAP}~\cite{zhang2020snap} is a sparse CNN accelerator where the index matching occurs before the element multiplication. To find index-matched data in parallel, the architecture of SNAP includes a large comparator array shared by multiple PEs. After finding the matched indices among the weight and the input, the information about the matched pair is encoded and transferred into each PE.
% To benefit from the encoded data, each PE includes decoder and address computation unit as well.
The decoders of the PEs find the pairs of the weight and input, and multipliers perform element multiplication using the decoded index data.
% By using the address computation with the indices, the partial products for the same output elements are accumulated with the previous partial sums.
The matched non-zero input/weight pairs are computed in the multiplier array, and computed multiplication results are written back to the psum buffer.

{\bfseries SCNN}~\cite{parashar2017scnn} is an Cartesian product-based accelerator for the sparse CNN. To exploit the Cartesian characteristic, each PE of SCNN includes a coordinate computation unit as an index matching unit. Cartesian product approach first performs matrix multiplications between all elements in the weight vector and the input vector.
% The accumulation of the partial results is the final output of the matrix multiplication. For the sparse neural networks, the multiplication between sparse vectors makes sparse partial results. To find the output position to add the partial product, the coordinate computation unit calculates the output index instead of matching non-zero input/weight pairs as in other sparse matrix multiplication accelerators. The partial products are accumulated in the psum buffer depending on the output index.
Then, the multiplication results are sent to the backend psum buffers using the index number generated from the coordinate computation unit.
% Because of the Cartesian computing method, the partial sum buffer needs a size of the output matrix at least.
To implement the Cartesian product approach, the psum buffer must be larger than the size of the dense output matrix to reduce the writeback traffic congestion.
%The additional units to find output index increase a chip-area and energy consumption.

%{\bfseries SNAP}\cite{zhang2020snap} is an sparse CNN accelerator that the index matching occurs before the element multiplication. To find a index-matched data in parallel, the architecture of SNAP includes a large comparator array shared by some PEs. After finding the matched indices among the weight and the input, the information about the matched pair is encoded and transferred into each PE. To benefit from the encoded data, each PE includes decoder and address computation unit as well. The decoders of the PEs figure out the pairs of the weight and input, and multipliers perform element multiplication according to the decoded information. By using the address computation with the indices, the partial products for the same output elements are accumulated with the previous partial sums.

{\bfseries SIGMA}~\cite{qin2020sigma} introduces a general sparse matrix multiplication accelerator that supports arbitrary shape and sparsity using unique interconnect networks. In this design, the non-zero elements of one sparse matrix are mapped onto the PEs, and a distribution network allocates non-zero elements from another sparse matrix to the PEs with matching non-zero elements for multiplication. Then, a reduction network produces multiple outputs by flexibly accumulating corresponding partial sums.
% Consequently, SIGMA achieves high PE utilization by broadcasting the input vector.
However, since SIGMA is based on a bitmap compression format, the index-matching logic needs to check a larger number of elements, including zeros, than the number of PEs in a single cycle to maintain computational throughput. This necessitates intricate routing logic to handle the collection of an arbitrary number of potential matching cases in a clock cycle.
% Additionally, information about an arbitrary number of non-zero pairs must be written to control the interconnect networks.
%However, the routing logic is significantly complex to collect an arbitrary number of possible matching cases in a clock cycle, and hence it is difficult to implement such a complex router logic.
As a result, the limited throughput in finding non-zero pairs results in a significantly longer time than the computation time, leading to an overall decrease in performance.

%In contrast to the aforementioned accelerators, 
{\bfseries S2TA}~\cite{liu2022s2ta} is a sparse CNN accelerator that leverages structured sparsity. Through the trade-off in accuracy resulting from the structured sparsity, it adopts a systolic array architecture similar to the dense neural network accelerator. Despite its relatively simple architecture, for the identification of non-zero pairs during multiplication, the data transferred between PEs is in the form of a vector with multiple non-zero elements. This is distinct from the data movement involving single element in the dense neural network accelerators. Within each PE, a multiplexer selects a non-zero pair for computation by comparing indices of these vectors.

Although many hardware accelerators tailored to the sparse neural networks have been introduced, such an approach leads to significant area/power overheads (Fig. \ref{prev_arch}b) due to the index-matching unit to match input/weight pairs, FIFOs and large buffers to balance load between PEs.
For example, in the area breakdown of the SCNN paper, the scatter network and FIFO occupy 3.75$\times$ higher area than the multiplier array, which reduces the computing density (throughput/area). Therefore, our Sparse-on-dense shows higher throughput/area than the baseline sparse hardware by up to 11.9$\times$, which will be discussed in Section IV.
Because the number of matched input/weight pairs the PEs have is irregular, the number of computations differs between the PEs.
When PEs continuously compute a larger number of output activations/psums, the idle time due to the irregular number of non-zeros for the PEs becomes smaller thereby increasing the average MAC utilization.
Therefore, the buffers in the PEs need to have large capacities to keep as many input/weight/psums as possible. 

As summarized in Table \ref{table_sparsehw}, the STA and S2TA used the structured sparsity. The accelerators with structured sparsity show high area and energy efficiencies, but such an approach leads to lower accuracy than unstructured pruning scheme due to the constraint of weight pruning pattern. So, our work focuses on the acceleration of unstructured sparsity-based neural networks with small area overhead. Meanwhile, our design can also compute the neural networks with structured sparsity by skipping decompression units, which will be explained in Section V.

\section{Sparse-on-Dense Architecture}
% Recently, a lot of DNN models have been developed with the increasing number of parameters. The developing trends of DNNs inspire the demand for the dense matrix based hardware that supports the sparse data. The sparsity is a powerful tool to reduce the burden of huge DNN models. However, the DNN models needs particular pruning recipes to maintain the accuracy. Because it is impossible that all DNN models have their own pruning method, the hardware accelerators needs generality to process the dense data. While the sparse matrix multiplication accelerator performs redundant operations for the dense data , the dense matrix based accelerators complete the computation with the simple operations. Therefore, the dense matrix multiplication hardware supporting the sparse data is required for the efficient processing of the various DNN models.

The main contribution of our work is to develop a hardware accelerator which can efficiently handle sparse neural networks by exploiting 1) the simplicity of dense matrix multiplication accelerators and 2) the reduction in the number of memory accesses on sparse matrix multiplication accelerators.
Furthermore, pruning techniques are powerful tools, but it takes time for researchers to develop the pruning recipes for each neural network. As a result, it is important to support not only sparse neural networks but also unpruned state-of-the-art networks to operate well on the hardware accelerators using simple dense matrix multiplication-optimized PE array. 

\subsection{Overall Architecture}

%The goal of Sparse-on-Dense is simple computation of sparse neural networks by integrating the advantages of both dense and sparse accelerators. 
Fig. \ref{sod_arch}a shows the architecture of Sparse-on-Dense which includes the global buffer for inputs/weights/psums, two decompression units, and a systolic PE array.
To take advantage of simplicity in dense matrix multiplication accelerators, we adopted the Google's TPU-style \cite{isca_tpu} systolic array in our PE design.
Meanwhile, the global buffer can store data in the sparse format for the memory saving. The decompression unit turns the sparse format into the dense format for the systolic PE array.
For dense network processing, the buffer holds the dense format data and directly feeds the PE array, bypassing the decompression unit (Fig. \ref{sod_arch}c). 
This detour allows the multiplication of the sparse format data and the dense format data (Fig. \ref{sod_arch}b) on Sparse-on-Dense. %need to use another representation
The accumulator completes matrix multiplication by adding the output of PE array with the previous output.

\subsection{Data Format and PE Array}

\subsubsection{Sparse Data Format in SRAM}

% To take advantage of the sparsity like sparse matrix multiplication accelerators do, Sparse-on-Dense allows to store the sparse data in the sparse format, which is the efficient method to store a sparse data. The support of the sparse format can reduce energy consumption for the data movement and the memory write operation of Sparse-on-Dense. As the compressed format includes the non-zero elements only, there is less data movement and memory usage for the single weight tile (Figure \ref{sparse_form_exp}). The less usage of the memory gives another chance to reduce the data movement.

Because our Sparse-on-Dense stores the non-zero values in the compressed format (Fig. \ref{sparse_form_exp}c), it can reduce the memory footprint compared to the case in which the dense format data is stored in the on-chip memory (Fig. \ref{sparse_form_exp}b).

When the size of the input tensor is larger than the capacity of on-chip memory, the large input tensor is typically tiled into multiple small tensors.
Then, the tiled input tensors are uploaded to the on-chip memory and participate in the matrix multiplication.
It is widely known that if the tiled tensor size is larger, the number of on-chip reuses for the inputs/weights increases.
By using the sparse input/weight matrices, the effective number of inputs/weights stored in the on-chip memory is much increased compared to the case using dense matrices.
As a result, the multiplication with the sparse matrices shows much larger on-chip reusability, reducing the number of data communication with the external memory,
% When the size of the input data is larger than the size of the on-chip memory, the same data need to be fetched repetitively.  With the sparse format, the chance for a data to stay in the on-chip memory increases.
%the less memory usage for the same data gives more capacity to store the other data. 
% If the entire data in the sparse format is smaller than the buffer size, but the dense format data is larger than the buffer size, the sparse data can be reused many times while the dense format data is re-fetched. 
% Sparse-on-Dense has the same memory access pattern as the conventional sparse neural network accelerators, and hence has the smaller number of data movement to external memory
which results in the smaller energy consumption for the matrix multiplication than the dense neural network accelerators.
%like sparse neural network accelerators. %how to describe simply

%The indices and values of the non-zero elements are stored in the sparse format.
%Generally, the computation of the sparse format data requires a special hardware architecture mitigating irregularity caused by the random appearance of the data. On the contrary, the dense neural network hardware accelerators using regular patter of the computation need to store the sparse data in the dense format including zeros. 

%The sparse format reduce the amount of data movement in two ways. First, the amount of data in the sparse format is basically smaller than the dense format for the sparse data as the compressed format includes the non-zero elements only. Furthermore, it is possible to reduce the number of tiles fetched during the multiplication. When the size of the data is larger than the size of the on-chip memory, the same data could be fetched several times. With the sparse format, the number of the tile that can be kept in the buffer is larger than with the dense format. If the entire data in the sparse format is smaller than the buffer size and the dense format data does not, the sparse data can be reused while the dense format data is re-fetched. As a result, the reduction of the data movement of Sparse-on-Dense leads the less energy consumption for the matrix multiplication like sparse neural network accelerators.
% better to add an example that shows the sparse format does not need re-fetch?

\subsubsection{Dense Matrix Multiplication in PE Array}
% Although the sparse format data is stored on the chip, Sparse-on-Dense performs matrix multiplication using the dense neural network accelerator's method. By adopting the systolic datafolw\cite{isca_tpu} using the dense format data, Sparse-on-Dense takes advantages of a simple hardware architecture. As the multiplication occurs without the additional operation, such as index matching, each PE consists of a MAC unit and small buffers for double-buffered weight, activation, and partial sum.  This simple hardware architecture leads less chip-area and less power consumption than PE arrays of the sparse neural network accelerators. Threfore, Sparse-on-Dense can have more PEs than the sparse processing hardware for the same area.

Previous sparse neural network accelerators obtain high PE utilization with complex logic including the index-matching unit, large buffers, and FIFOs. However, the additional components incur significant overheads in area and power consumption. Instead, we mapped the sparse matrix onto the dense matrix multiplication-based PE array \cite{isca_tpu}. Although such an approach decreases the PE utilization, the simple PE array shows higher performance/area than sparse matrix multiplication-based PE array. The detailed evaluation will be presented in Section~\ref{section_result}.

% \subsubsection{Decompression between SRAM and PE array}
\subsubsection{Decompression Unit}
As Sparse-on-Dense uses the dense-PE, it needs an on-chip decompression unit to transform the sparse format data into the dense format.
If the decompression unit is located before the global buffer, the global buffer needs to store larger dense format data.
Instead, to maximize the advantage from the compactness of the sparse format data, we place the decompression units in the stage between the global buffer and the PE array (Fig. \ref{sod_arch}). 
The decompression unit consists of a pointer buffer, a non-zero index/value buffer, a subtractor, an element selection unit, a dense mapping unit, and a dense format buffer.
% The size of the non-zero buffer receiving the compressed data from the global buffer has the same size with the bandwidth of the global buffer to maximize the utilization of the bandwidth.
The size of the non-zero buffer receiving the compressed data from the global buffer is set to fully utilize the bandwidth between the global buffer and the PE array.
% The number of elements in the dense format buffer issuing the dense format data to the PE array can store the same number of elements with the number of rows/columns of the PE array. The decompression operation can be separated into two parts, obtaining the relative index, which means the number of non-zero elements in a single column of the matrix, and making dense format data from the sparse format data. The attainment of how many elements to feed is essential to select the necessary data among the sparse data coming from the global buffers. The value of the selected data locates at the position of the dense vector. 

% Fig. \ref{decomp_step} shows the order of the operations to decompress the CSC format. The decompression starts from fetching the pointer to the pointer buffer of the decompression unit (Fig. \ref{decomp_step} 1)). After the pointer fetch is over, the number of nonzero elements in each column is obtained by subtracting two sequential pointer values. Meanwhile, the indices and values of the non-zero elements are transferred to the non-zero buffer from the global buffer (Fig. \ref{decomp_step} 2)). 
Fig. \ref{decomp_step} shows the processing steps to decompress the CSC format. First, the pointer values are loaded from the global buffer and stored in the pointer buffer of the decompression unit (Step 1). After the pointer fetch is over, the number of non-zero elements in each column is obtained by subtracting two sequential pointer values. Meanwhile, the indices and values of the non-zero elements are transferred to the non-zero buffer from the global buffer (Step 2).
% As the non-zero buffer takes as many non-zero data as possible, the elements fetched simultaneously can be located in the different columns of the original matrix. As a result, the decompression unit selects items that have the same column index as the original matrix based on the subtraction result of the pointers (Step 3).
Because the non-zero buffer takes as much non-zero data as possible, the simultaneously fetched non-zero elements in a non-zero buffer row can be located in the different columns of the original matrix. Therefore, the decompression unit sends the items to the corresponding index of the dense format buffer depending on the original matrix column index obtained from the subtraction result of the pointers (Step 3/4).
% The indices of the selected elements are the indices in the dense format buffer (Step 4).
Finally, the decompressed data is fed to the PE array (Step 5).

% \vspace{\baselineskip}

% In short, Sparse-on-Dense takes advantages in the next aspects. Compared to the sparse neural network accelerators, Sparse-on-Dense has low hardware cost maintaining the small memory requirements. Considering the dense neural network accelerator, Sparse-on-Dense reduce the data movement with a little overhead in area and power consumption.

\begin{figure}[t]
  \centering
  \includegraphics[width=\linewidth]{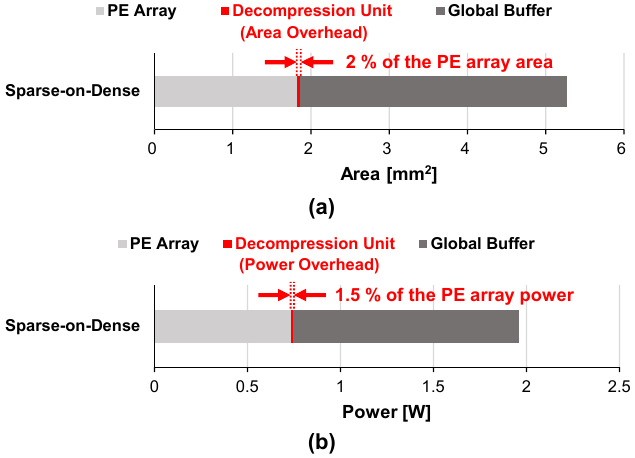}
  \caption{The area and power breakdown of Sparse-on-Dense used for the dense neural network comparison. PE array has 4K PEs and 2 MB SRAM is implemented.}
  %\Description{Area breakdown shows the area overhead.}
  \label{area_breakdown}
\end{figure}

\section{Experimental Results}
\label{section_result}
\subsection{Experiment Setup} %what to say about setting
% We compare Sparse-on-Dense with the dataflow of the state-of-the-art neural network accelerators. To evaluate the efficiency for the various density, the estimation was performed by sweeping the density from 10\% to 100\%. 
In this Section, we evaluate the Sparse-on-Dense architecture. We compare the Sparse-on-Dense with 2 baseline scenarios including  1) the dense matrix multiplication acceleration \cite{isca_tpu} and 2) mapping the sparse matrix multiplication to the sparsity-aware neural network accelerators \cite{han2017ese,parashar2017scnn,zhang2020snap,qin2020sigma}. For a fair comparison, we adopted the original configuration used in each of previous works, with the exception of the SIGMA \cite{qin2020sigma}. Since the number of logic gates described in \cite{qin2020sigma} (1024) is smaller than the number of PEs (128$\times$128), we employed 16384 AND gates to find the matched non-zero pairs, accompanied by an 8-port SRAM to store non-zero pair information. To evaluate the efficiency for the various density, the estimation was performed by sweeping the density from 0.1-to-1.0. 
%Furthermore, to show the validity of Sparse-on-Dense on the actual workloads, we evaluated several network models such as AlexNet, VGG-16, and BERT (Table \ref{real_sparsity}). 
%%the evaluation for the sparse neural networks (Table \ref{real_sparsity}) was also performed.
For the analysis of throughput, we evaluated the throughput per compute area.
For the energy-efficiency, the system-level energy consumption including DRAM, the global SRAM buffer and the PE array was analyzed.

%For the fair comparison, we matched hardware configuration, such as the number of MACs and the size of memory, with the baseline architectures. 
% We created the cycle-accurate simulator for the performance evaluation.
% use published sigma code on github
% For the analysis of the energy consumption and area, we designed the accelerators in Verilog HDL and synthesized in a 28nm CMOS technology using Synopsys Design Compiler.
For the analysis of the energy consumption and area, we implemented all the baselines and our design in Verilog HDL and synthesized in a 28nm CMOS technology using Synopsys Design Compiler.
The Verilog implementation for the SIGMA~\cite{qin2020sigma} is available on github, so the Verilog model was directly adopted in the evaluation.

\subsection{Comparison with Dense NN Hardware} %pause

\begin{figure}[t]
  \centering
  \includegraphics[width=\linewidth]{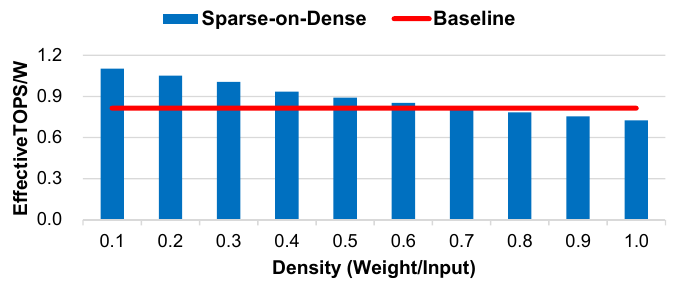}
  \caption{The energy efficiency of baseline dense hardware and Sparse-on-Dense for various densities. Note that the dense baseline always receives the dense-format data from external memory regardless of the density.}
  %\Description{sparse data reduce energy.}
  \label{dense_eng_eff}
\end{figure}

%how to mention the configuration of the dense NN baseline?, why weight sparsity, not dual-side sparsity
We first compared Sparse-on-Dense with the dense NN accelerator \cite{isca_tpu}. Both accelerators have the 4K systolic PE array and 2 MB global buffer, and they operate on 500 MHz. We used 16-bit precision for the inputs/weights and 8-bit for the indices of non-zero values.
% Because the PE array architecture and dataflow of Sparse-on-Dense are identical with the baseline accelerator, both accelerators can deal with the same number of MAC operation for the same time. The difference comes from the decompression unit.
Considering that we adopted the dataflow of the baseline accelerator in our Sparse-on-Dense, both accelerators can deal with the same number of MAC operations for the same time period. The difference between two designs mainly comes from the decompression unit.

\begin{table}[t]
\centering
\caption{Comparison of the throughput/area  $[$Effective TOPS/$mm$\textsuperscript{2}$]$ for the dense matrix based accelerator and Sparse-on-Dense when processing the dense data (density = 1.0).}

\begin{tabular}{|c|c|c|}
\hline
 Architecture & Logic & Logic + SRAM \\
\hline
Baseline & 0.956 & 0.430 \\
Sparse-on-Dense & 0.946 & 0.428 \\
\hline
\end{tabular}

\label{dense_area}
\end{table}

The area of the decompression unit is about 2\% of the area of the 4K PE array (Fig. \ref{area_breakdown}). Considering the entire chip area including memory, the area overhead of the decompression unit becomes smaller. Table \ref{dense_area} shows the throughput/area of the dense neural network accelerator and Sparse-on-Dense.
% Thanks to the small cost of the decompression unit, the logic circuits of Sparse-on-Dense has similar TOPS/area with the baseline architecture with about 1\% of degradation. The gap of the throughput\/area between two architectures becomes negligible when including the memory.
Thanks to the small area of the decompression unit, the logic circuits of Sparse-on-Dense has similar TOPS/area to the baseline architecture (about 1\% of degradation). 
%The gap of the throughput\/area between two architectures becomes negligible when including the memory area.

% The power evaluation is shown in Fig. \ref{dense_eng_eff}. When the weight data is sparse under 50\% of the density, the energy efficiency of Sparse-on-Dense is better than that of the baseline architecture. The additional power consumption of the decompression unit is small enough to benefit from the less memory requirements.  The result shows that the trade-off between area and energy for the decompression unit allows efficient computation for the sparse neural networks. %need to add why >0.5 sod worse and why similar at density=1
The energy efficiency analysis is shown in Fig. \ref{dense_eng_eff}. In this evaluation, we assume that the dense baseline always receives the dense format data from memory regardless of the sparsity level for Sparse-on-Dense.
% Sparse-on-Dense receives the dense format data when density > 0.7 because receiving sparse format data for the density that is higher than 0.7 shows the worse energy efficiency than the density = 1 case.
When the weight data is sparse (density $<$ 0.7), the energy efficiency of Sparse-on-Dense is better than that of the baseline architecture.
On the other hand, if the weight data is dense (density $\ge$ 0.7), the baseline shows higher energy-efficiency than that of our Sparse-on-Dense scheme.
The additional power consumption of the decompression unit is small enough to benefit from the smaller memory requirements. The result shows that the trade-off between area and energy for the decompression unit allows efficient computation for the sparse neural networks.

\subsection{Comparison with Sparse NN Hardware} 
\begin{figure}[t]
  \centering
  \includegraphics[width=\linewidth]{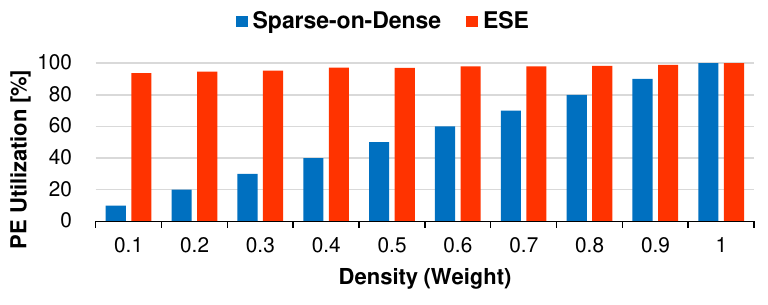}
  \caption{Comparison of utilization with ESE hardware \cite{han2017ese}.}
  \label{ese_util}
\end{figure}

\begin{figure}[t]
  \centering
  \includegraphics[width=\linewidth]{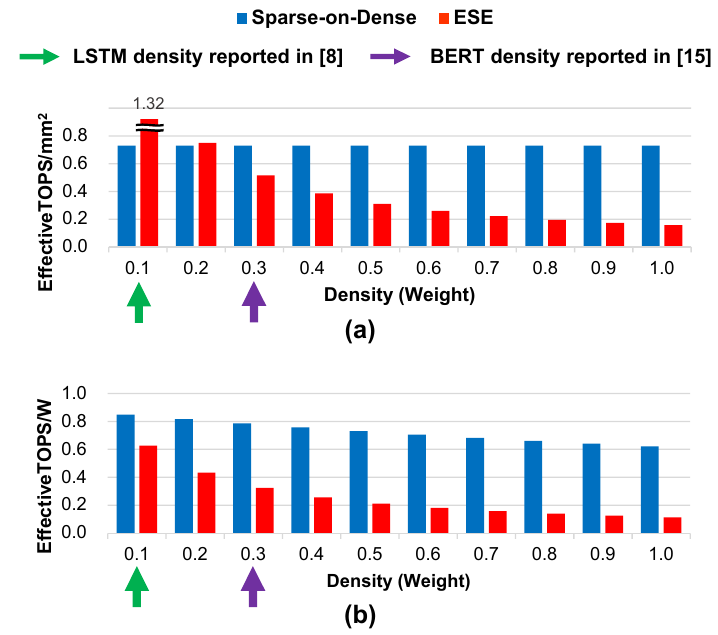}
  \caption{Comparison of throughput/area and energy-efficiency between the proposed design and ESE. The density of the sparse LSTM is 0.1 \cite{han2017ese}. BERT (SQuAD) requires at least 30\% of density to maintain accuracy\cite{sanh2020movement}.}
  %\Description{The evaluation result shows better efficiency.}
  \label{ese_comp}
\end{figure}

In this Section, we compared Sparse-on-Dense with the sparse neural network accelerators for the various density.
% Because Sparse-on-Dense takes advantage of dense matrix multiplication hardware, we evaluated the performance per area and energy efficiency of the logic circuits and the system including SRAM. 
% For the fair comparison, the evaluation is performed using Sparse-on-Dense that has the same configuration, including operation frequency, bit precision, the number of MACs and the size of memory, with the baseline accelerators.
For the fair comparison, we used the same configurations, such as operation frequency, bit precision, the number of MACs and the size of buffers, for the Sparse-on-Dense and the baselines.
% Considering that the sparse neural network requires less computations to complete the matrix multiplication, effective TOPS is adopted as the performance benchmark. The effective TOPS, obtained by dividing TOPS for the density, indicates the performance of the dense operations that makes the same output with the sparse operations.
For the performance analysis, the effective throughput, which is the throughput divided by matrix density, was compared between dense matrix multiplication hardware and sparse matrix multiplication hardware.
 
\subsubsection{Sparse Weight and Dense Input}  Fig. \ref{ese_comp} shows the comparison between Sparse-on-Dense and ESE~\cite{han2017ese}. Because ESE is designed for the sparse LSTM, the evaluation is performed for the multiplication between the sparse weight and the dense input. This combination is frequently used for the natural language processing (NLP) models, including LSTM and BERT. 

% The comparison of the area is in Fig. \ref{ese_comp}a and Fig. \ref{ese_comp}b. When the weight density is 10\%, the logic circuits of ESE has higher throughput for the same area thanks to high PE utilization. When the number of non-zero elements in the weight matrix increases, Sparse-on-Dense shows higher throughput from the same area on the logic. In other words, the area overhead of complex PE architecture in ESE rises to the surface. This is because the improvement of PE utilization is small even though there are more elements to compute. When including SRAM, the large size of SRAM implemented to ESE changes the trend of the evaluation. If the SRAM becomes small, the performance related to the area would follow the logic circuit's.

We first compared the multiplier array utilization between two hardware accelerators (Fig. \ref{ese_util}).
In our Sparse-on-Dense accelerator, sparse matrices are decompressed before being sent to the systolic array. Considering that the systolic array computes the dense matrix multiplication, utilization is lower than the ESE hardware in all density cases.
However, the ESE had to use a large design area to obtain the high multiplier array utilization. 
For a fair comparison, we analyzed the throughput/area of these accelerators.
The throughput/area of Sparse-on-Dense was compared to the ESE baseline in Fig. \ref{ese_comp}a.
Note that the throughput/area of Sparse-on-Dense remains constant for all density values because the number of active PEs and the density scale together. 
When the weight density is 0.1, the ESE shows higher (1.8$\times$) throughput for the same area because the MAC utilization of the ESE is much higher than ours.
However, the ESE requires large-sized FIFOs, buffers, and the index-matching unit to maintain high MAC utilization.
Because the ESE requires a much larger logic area than ours to compute the same number of MAC computations in almost all densities except for density$<$0.3, the Sparse-on-Dense shows better throughput/area than the ESE when density$>$0.2.
%The energy-efficiency of the ESE baseline is better than ours over the all density ranges.
The energy-efficiency of our design is higher than that of the ESE for all the densities.
% when density is smaller than 0.4 (Fig. \ref{ese_comp}b). However, the Sparse-on-Dense shows better energy-efficiency than the ESE in the higher densities than 0.4.
At the typical workload densities, the Sparse-on-Dense shows better throughput/area and energy-efficiency by 0.8-1.4$\times$ and 1.4-2.4$\times$ than the ESE baseline, respectively.

% Fig. \ref{ese_comp}c and Fig. \ref{ese_comp} are the energy efficiency. For overall density, Sparse-on-Dense has better energy-efficiency than ESE because of the low hardware complexity. Different from the simple hardware using systolic dataflow repeating MAC and saving, ESE performs special operations to obtain high utilization with sparse data. Therefore, the complexity of circuits increases, making degradation of the energy efficiency. 

\begin{figure}[t]
  \centering
  \includegraphics[width=\linewidth, scale=1.5]{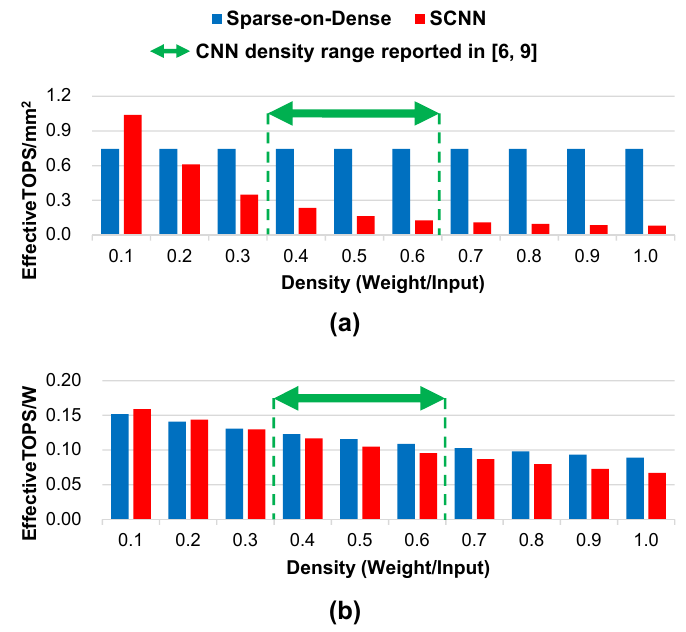}
  \caption{Comparison of throughput/area and energy-efficiency between the proposed design and SCNN~\cite{parashar2017scnn}.} 
  %The density range of sparse CNN is 0.4-0.6 \cite{han2015deep, zhang2020snap}}
  %\Description{The evaluation result shows better efficiency for both the sparse range and dense range.}
  \label{scnn_comp}
\end{figure}

\begin{figure}[t]
% \vspace{-5mm}
  \centering
  \includegraphics[width=\linewidth]{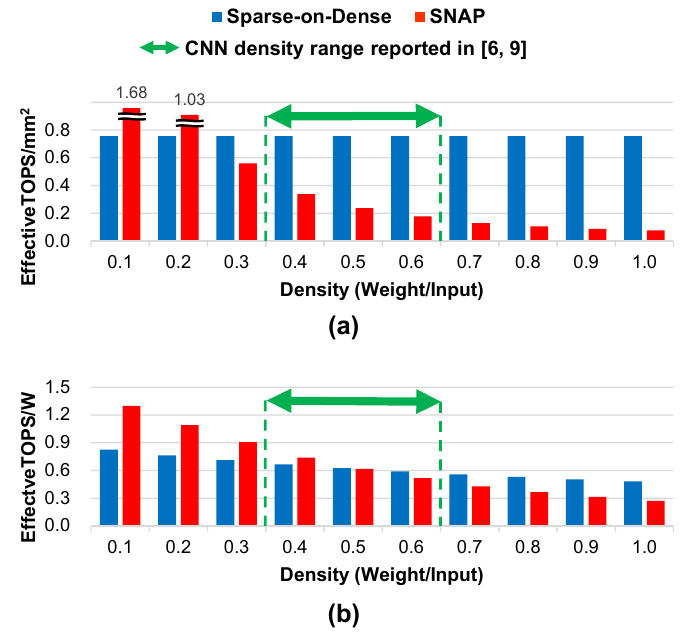}
  \caption{Comparison of throughput/area and energy-efficiency between the proposed design and SNAP~\cite{zhang2020snap}.} 
  %The actual sparse workloads have 0.4-0.6 of density\cite{han2015deep,zhang2020snap}.}
  %\Description{The evaluation result shows better efficiency for the density of the real applications.}
%   \vspace{-5mm}
  \label{snap_comp}
\end{figure}

\begin{figure}[t]
% \vspace{-5mm}
  \centering
  \includegraphics[width=\linewidth]{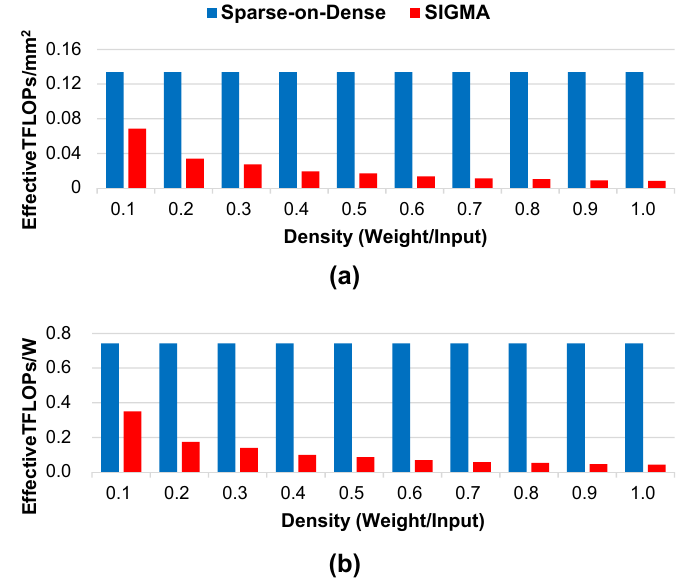}
  \caption{Comparison of throughput/area and energy-efficiency between the proposed design and SIGMA~\cite{qin2020sigma}.} 
%   \vspace{-5mm}
  \label{sigma_comp}
\end{figure}

\subsubsection{Sparse Weight and Sparse Input}
\label{Sp_W_I}
% We compared two types of sparse CNN accelerator with Sparse-on-Dense.
We also compared our Sparse-on-Dense with two-sided sparsity-aware CNN accelerators \cite{parashar2017scnn,zhang2020snap}.
% In the case of sparse CNN, the input sparsity caused by ReLU activation function makes the complex index matching operation between two sparse data. The sparse CNN accelerators can be separated into two types depending on whether the index matching is performed before multiplication or not. 

% Fig. \ref{scnn_comp}a and Fig. \ref{scnn_comp} demonstrate the comparison with the Cartesian-product-based approach, SCNN\cite{parashar2017scnn}. In Fig. \ref{scnn_comp}a, Sparse-on-Dense shows the higher logic performance per area except for 10\% density. The gap between two architecture expands as the density increases. This is because scattering output of the multiplier array disturbs improvement of the PE utilization, even though the multipliers can be used as many as possible. Different from SCNN, Sparse-on-Dense maintains the throughput for the unit area because of the regular access patter for the output. Furthermore, the area of scattering crossbar is 3.25x of the multiplier array\cite{parashar2017scnn}. The additional circuit to support sparsity becomes the reason of the low performance. Even though the memory area alleviate the index matching overhead of ESE, Sparse-on-Dense shows better result over 50\%.

Fig. \ref{scnn_comp}a shows the comparison of throughput/area between the SCNN~\cite{parashar2017scnn} and our design. Different from the ESE design where index-matching unit is necessary to match non-zero input/weight pairs, Cartesian product does not require such an index-matching logic for the non-zero inputs. However, the SCNN needs a complex scatter network and a large-sized psum buffer to store generated psums in the target address of the buffer, which leads to significant area/power overhead. Furthermore, the gap between two architecture increases as the density increases. This is because the scatter network in the backend of the multiplier array shows traffic congestion. In contrast, Sparse-on-Dense maintains the high throughput for the unit area thanks to the regular address pattern for the output. At the typical workload densities, our Sparse-on-Dense shows higher throughput/area by 3.1-5.8$\times$ and higher energy-efficiency by 1.0-1.1$\times$ over the SCNN baseline design.

The PE array of SNAP~\cite{zhang2020snap} uses an index-matching unit, FIFOs, a large-sized input/weight/psum buffers to maximize the load balance and multiplier array utilization similar to the ESE case. When the density is extremely low, the throughput/area and energy-efficiency of the SNAP is better than ours thanks to the index-matching method. However, due to the large overheads in area and power to maintain high utilization, our design shows higher throughput/area and energy-efficiency than the SNAP when the density is not significantly low.
At the typical workload densities, our Sparse-on-Dense shows better throughput/area and energy-efficiency by 2.2-4.2$\times$ and 0.9-1.1$\times$ than those of the SNAP, respectively.

The computing units of SIGMA~\cite{qin2020sigma} are optimized for sparse matrix multiplication through its interconnect network. However, the reduction network in SIGMA incorporates the same number of buffers as the number of PEs in each level of its reduction tree structure, enabling arbitrary output paths with pipelining support, which largely increases logic area.
% Additionally, the limited throughput of the index matching logic reduces their computing ability even with the employment of the same number of logic gates as PEs. 
Furthermore, it is difficult to implement the complex router logic to collect an arbitrary number of possible matching cases in a clock cycle as explained in Section \ref{prelim_sparse}.
The inefficiency of the control flow significantly diminishes the overall performance of the architecture, resulting in reduced efficiency in terms of both area and energy (Fig. \ref{sigma_comp}). On the other hand, our design does not need additional index-matching process for a large number of elements, although it requires decompression for a small number of non-zero elements. Furthermore, the systolic array in our design features the straightforward and regularized output path. Consequently, for all typical workload densities mentioned previously, our Sparse-on-Dense achieves higher throughput/area by 1.9-9.7$\times$ and higher energy-efficiency by 2.1-10.1$\times$ compared to SIGMA.

%\subsection{Realistic Density of the Sparse Applications} 
\subsection{Evaluation using Real Benchmark Datasets}%or subsection?
\label{eval_datasets}
\begin{figure*}[t]
  \centering
  \includegraphics[width=\linewidth]{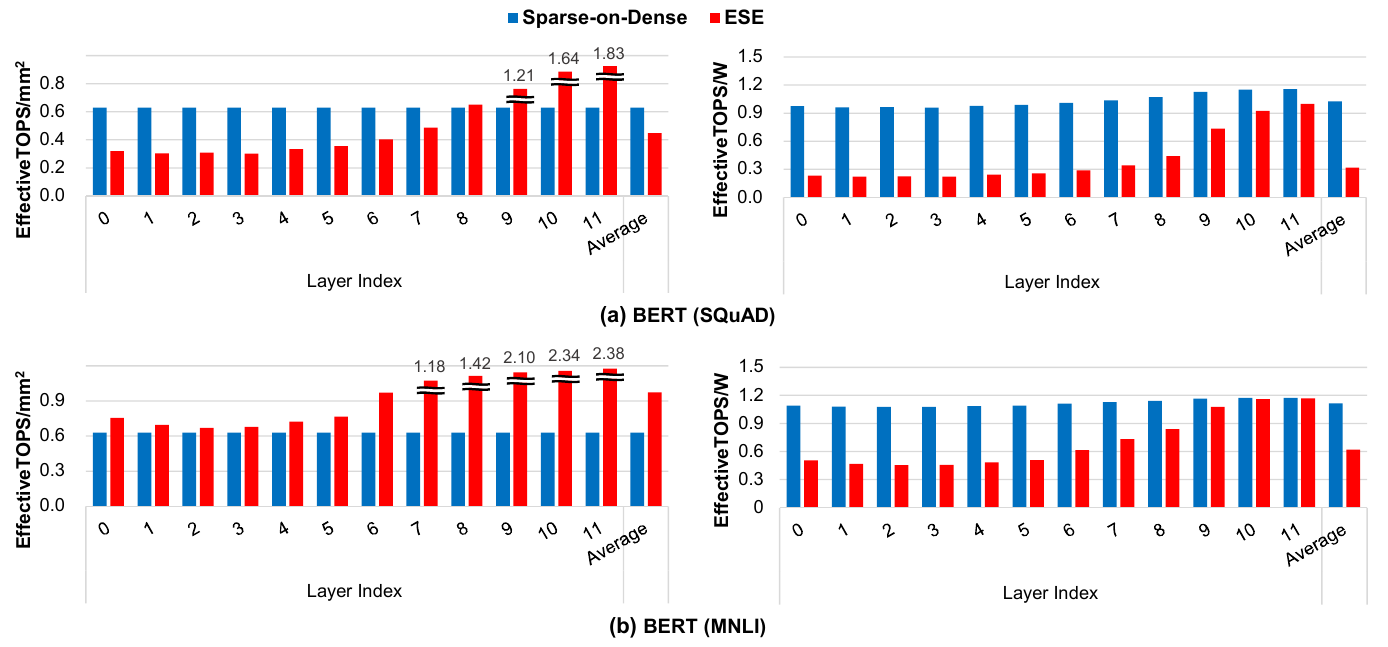}
  \vspace{-5mm}
  \caption{Layer-wise comparison of throughput/area and energy-efficiency between the Sparse-on-Dense and ESE~\cite{han2017ese} for the pruned BERT model trained for (a) SQuAD dataset and (b) MNLI dataset.}
  \vspace{-5mm}
  %\Description{The evaluation result shows better efficiency for BERT-SQuAD dataset.}
  \label{ese_comp_real}
\end{figure*}

\begin{figure*}[t]
  \centering
  \includegraphics[width=\linewidth]{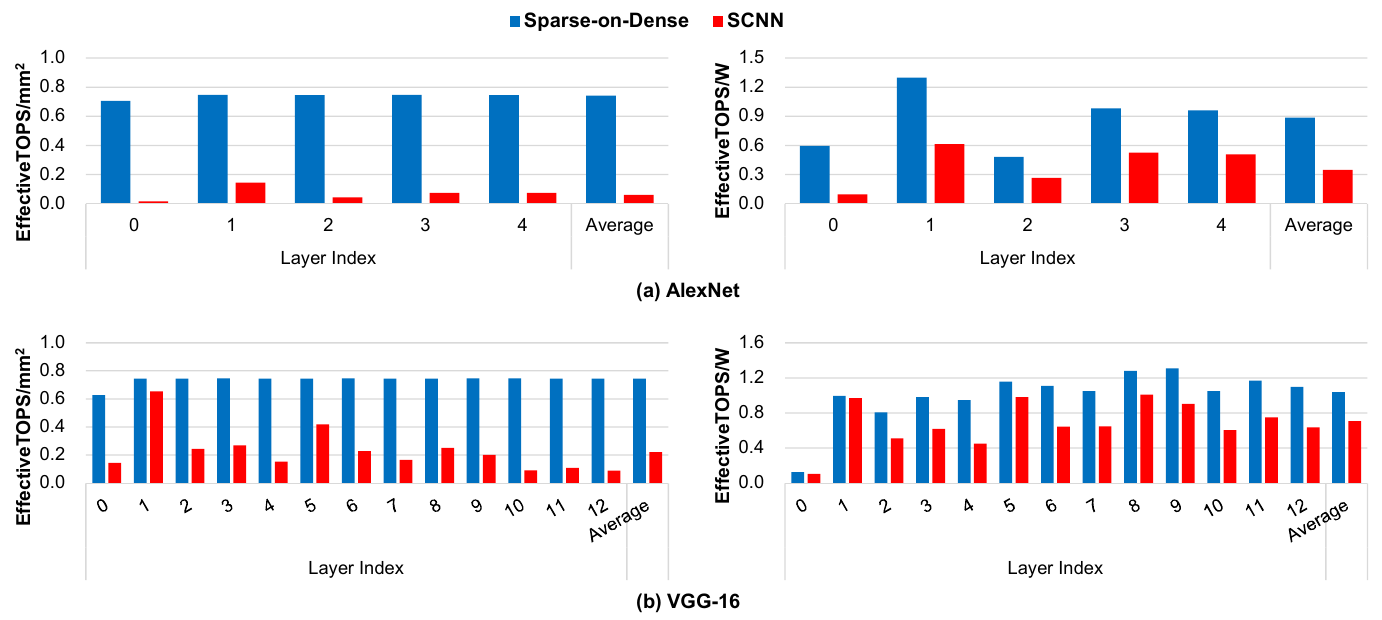}
  \vspace{-5mm}
  \caption{Layer-wise comparison of throughput/area and energy-efficiency between the Sparse-on-Dense and SCNN~\cite{parashar2017scnn} for convolutional layers of the pruned (a) AlexNet and (b) VGG-16. Note that, unlike Sparse-on-Dense, SCNN cannot deal with the fully connected layers.}
  \vspace{-5mm}
  %\Description{The evaluation result shows better efficiency than SCNN for the pruned AlexNet.}
  \label{scnn_comp_real}
\end{figure*}

\begin{figure*}[t]
  \centering
  \includegraphics[width=\linewidth]{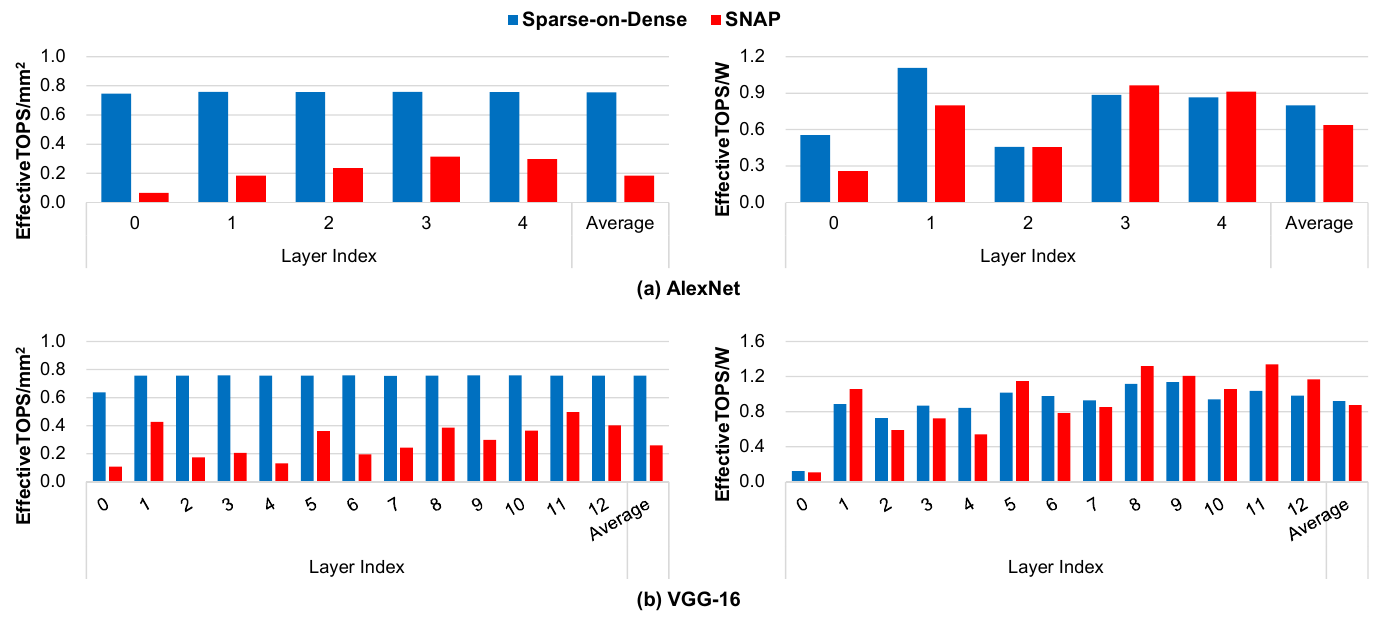}
  \vspace{-5mm}
  \caption{Layer-wise comparison of throughput/area and energy-efficiency between the Sparse-on-Dense and SNAP~\cite{zhang2020snap} for convolutional layers of the pruned (a) AlexNet and (b) VGG-16}
  \vspace{-5mm}
  %\Description{The evaluation result shows better efficiency than SNAP for the pruned AlexNet.}
  \label{snap_comp_real}
\end{figure*}

To show the validity of the proposed Sparse-on-Dense design in more realistic conditions, we further evaluated several sparse neural network models on the actual benchmark datasets: ImageNet dataset for AlexNet and VGG-16, SQuAD and MNLI datasets for BERT~\cite{devlin2018bert}.  
We pruned the AlexNet and VGG-16 models following the methods in \cite{han2015learning}, and pruned the BERT models using the methods suggested in \cite{sanh2020movement}. We reproduced similar weight/input density and accuracy to the values that were reported in \cite{han2015learning,sanh2020movement} (Table \ref{real_sparsity}).
%To verify the practicality of Sparse-on-Dense on the real applications, we surveyed the density of the actual sparse neural networks. In the case of NLP, the weight density of LSTM can be reduced to 0.1 according to \cite{han2017ese}. 
%BERT needs at least 30\% of the weight parameters to maintain the accuracy \cite{sanh2020movement}. 
%Even at the extremely low density(=0.1), the Sparse-on-Dense shows better energy-efficiency than the Dense NN accelerator and ESE baselines. However, the throughput/area of the ESE is much higher than our design at the density level.

%%real sparse network table - will be edited later
\begin{table}[t]
\centering
\footnotesize
\caption{ Density of the sparse DNN models used for the experiment.}

\begin{tabular}{ |c||c|c|c|c| }
\hline
 \multirow{2}{*} {Network} & Average  & Average & Accuracy & Pruning \\
  & weight density & input density & degradation & method  \\
\hline
%AlexNet CONV & 0.37 & 0.72 & 0\% & \cite{han2015learning}\\
%VGG-16 CONV & 0.32 & 0.65 & 0\% & \cite{han2015learning}\\
%ResNet50 CONV & 0.22  & 0.61 & 0\% & \cite{admm_pruning} \\
%BERT (SQuAD) & 0.33 & 1.00 & \% & \cite{sanh2020movement}\\
%BERT (MNLI) & 0.12 & 1.00 & 3.1 \%& \cite{sanh2020movement} \\
AlexNet & \multirow{2}{*} {0.41} & \multirow{2}{*} {0.69} & \multirow{2}{*} {0\%} & \multirow{2}{*} {\cite{han2015learning}}\\
CONV & & & & \\
VGG-16 & \multirow{2}{*} {0.33} & \multirow{2}{*} {0.61} & \multirow{2}{*} {0\%} & \multirow{2}{*} {\cite{han2015learning}}\\
CONV & & & & \\
%ResNet50 & \multirow{2}{*} {0.22}  & \multirow{2}{*} {0.61} & \multirow{2}{*} {0\%} & \multirow{2}{*} {\cite{admm_pruning}} \\
%CONV & & & & \\
BERT & \multirow{2}{*} {0.33} & \multirow{2}{*} {1.00} & \multirow{2}{*} {3.2\%} & \multirow{2}{*} {\cite{sanh2020movement}}\\
(SQuAD) & & & & \\
BERT & \multirow{2}{*} {0.12} & \multirow{2}{*} {1.00} & \multirow{2}{*} {3.1\%} & \multirow{2}{*} {\cite{sanh2020movement}} \\
(MNLI) & & & & \\
\hline
\end{tabular}

\label{real_sparsity}
\end{table}

When pruning the BERT models, the density of the pruned networks varies depending on the target tasks \cite{sanh2020movement}. Fig. \ref{ese_comp_real} compares the effective throughput per area and energy-efficiency of Sparse-on-Dense and the baseline ESE \cite{han2017ese} for the BERT models on 2 different datasets. %which have different densities. 
When pruned for the SQuAD \cite{rajpurkar2016squad} dataset, the range of weight matrix densities are 0.04--0.5 and the average density is 0.33. 
%Except for the last three layers that have lower densities than 0.2, the performance of Sparse-on-Dense was 1.0--2.1$\times$ higher than ESE. For the entire layers, Sparse-on-Dense shows higher energy-efficiency with 1.2--4.3$\times$ improvement.
On average, the effective throughput per area and energy-efficiency of Sparse-on-Dense are 1.4$\times$ and 3.2$\times$ higher than ESE.

In case of MNLI dataset~\cite{wang2018glue}, BERT model can be pruned with lower weight matrix densities than the SQuAD case.
%Different from the SQuAD dataset, BERT model trained for the MNLI \cite{wang2018glue} dataset can be pruned with lower weight matrix densities.
The pruned weight density of each layer varied from 0.01 to 0.22 with the average of 0.13. Because the weight density of each layer is lower than or near 0.2, the average effective throughput per area of Sparse-on-Dense is lower than the baseline accelerator.
%in all layers, and the average average throughput follows the results. 
However, the average energy efficiency of Sparse-on-Dense is 1.8$\times$ higher than ESE. Furthermore, it is worthwhile to note that Sparse-on-Dense can deal with the dense matrix computation such as attention algorithm  that the ESE accelerator cannot deal with, so the Sparse-on-Dense has more flexibility to handle overall NN computing acceleration.

For evaluating the efficiency of the proposed Sparse-on-Dense design for sparse convolutional neural network computing, we compare the proposed design with 2 previous works: SCNN~\cite{parashar2017scnn} and SNAP~\cite{zhang2020snap}. 
Fig. \ref{scnn_comp_real} shows the performance comparison between Sparse-on-dense and SCNN for computing AlexNet and VGG-16 models on ImageNet dataset.
In the sparse AlexNet model, the range of input and weight matrix densities are 0.38-1.0 and 0.34-0.84, respectively. The average input and weight matrix densities are 0.69 and 0.41, respectively.
When compared to the SCNN, the Sparse-on-Dense showed 5.1-16.5$\times$ performance improvement except for the first layer. 
%This noticeably higher performance gain than the results from the analysis with density-sweeping (Fig.\ref{scnn_comp} in Section \ref{Sp_W_I}) comes from the fact that, stride=4 is used for the first layer of the AlexNet but Cartesian product-based computation in SCNN is inefficient to handle sparse matrix multiplication for stride > 1 case.
% So, the SCNN paper did not consider the stride>1 case. 
The first layer computation with stride=4 on SCNN shows extremely low PE utilization (18\%), while our design shows 79\% PE utilization rate for computing of the first layer of the AlexNet. As a result, the Sparse-on-Dense architecture shows the 11.9$\times$ improvement of the average effective throughput per area compared to the SCNN baseline.

In case of VGG-16 network model, the input/weight matrix density ranges are  0.31-1.0 (input) and 0.22-0.57 (weight), and the average densities are 0.61 (input) and 0.33 (weight), respectively.
%In the layer 3-4 of AlexNet, the matrix densities are relatively lower than the front part (layer 0-2) of the layers. Hence, SNAP shows slightly higher effective energy-efficiency than our Sparse-on-Dense in those layers. However, the front layers of the AlexNet account for the larger number of computations than the rest layers. As a result, our Sparse-on-Dense shows higher energy-efficiency by 1.26$\times$ than the SNAP architecture.
% input/weight matrix density ranges of the VGG-16 network are 0.31-1.0(input) and 0.22-0.57(weight), and the average densities are 0.61(input) and 0.33(weight).
The average throughput per area and the energy-efficiency of the Sparse-on-Dense are 3.3$\times$ and 1.5$\times$ higher than those of SCNN. The layer 1 has much lower matrix density than the other layers, so the SCNN shows comparable throughput/area to our design in layer 1 of VGG-16 network.

It is interesting to see that the relative advantage of Sparse-on-Dense over SCNN in terms of energy efficiency is much larger in real benchmark simulations (Fig. \ref{scnn_comp_real}) than in density-sweep analysis (Fig. \ref{scnn_comp}b).  
%Some readers may have a question with the energy simulation results of the SCNN, because there are some inconsistencies between the density-sweep graph (Fig. \ref{scnn_comp}b) and real benchmark cases (Fig. \ref{scnn_comp_real} - Effective TOPS/W).
%In the density-sweep graph, the SCNN shows the comparable energy-efficiency to our Sparse-on-Dense. On the other hand, real benchmark results of the SCNN indicate the energy-efficiency of the SCNN is much lower with AlexNet/VGG-16 cases.
The performance evaluation with various density cases used the spatial domain kernel size=1 with which the SCNN shows the best performance among the various kernel sizes. However, realistic neural network models including the AlexNet/VGG-16 typically use much larger kernel sizes than 1. With the larger kernel sizes, our Sparse-on-Dense shows better energy-efficiency than the SCNN by optimizing the psum reuses for a dot product computation better.
%As a result, the energy-efficiency of the accelerators varies depending on the neural network parameters (eg., spatial-domain and input/output channel sizes).

Due to the limitation in the SCNN-like Cartesian-product based sparse neural network accelerators, inner-product based sparse neural network accelerators such as SNAP are actively studied.
Therefore, we also compare the Sparse-on-Dense with SNAP for the AlexNet and VGG-16 models on ImageNet dataset (Fig.\ref{snap_comp_real}).
In the layer 3-4 of the AlexNet, the matrix densities are relatively low compared to the earlier layers (layer 0-2). Hence, SNAP shows slightly higher effective energy-efficiency than our Sparse-on-Dense in those layers. However, the earlier layers of the AlexNet account for the larger number of computations than the rest layers. As a result, the Sparse-on-Dense shows 1.26$\times$ higher energy-efficiency  than the SNAP.
In case of the VGG-16 model, the average energy-efficiency gain over SNAP is 5\%, which is smaller than AlexNet case.
Compared to AlexNet, there is a larger portion of layers in VGG-16 in which SNAP shows higher energy-efficiency due to relatively low matrix densities.

%\vspace{5mm}
\section{Discussion}
\subsection{Dense Neural Networks and Structured Sparsity}

In the Section \ref{eval_datasets}, we tried to make a quantitative comparison between the Sparse-on-Dense and previous works in terms of TOPS/$mm^2$ and TOPS/W. However, although it is not straightforward to quantify, the most important merit of the proposed Sparse-on-Dense design is that it can handle any unstructured sparse neural network computations as well as dense network with competitive performance and energy-efficiency.
The structured sparsity is also handled in the same manner as the dense neural network by skipping decompression units in the input/weight loading stages.
In contrast, previous sparse neural network accelerators typically had a design focus which targets a specific neural network with a certain sparsity.
We expect that this flexibility will become even more important in multi-tenant neural network computing for which a single hardware is used to compute various types of neural networks.

\subsection{Power-gating for Additional Power Saving}
In some neural network layers, the MAC array is typically under-utilized due to the small number of channels, which increases the energy consumption. To maximize energy-efficiency, we can also consider the power-gating scheme. In this case, our systolic array can be divided into multiple voltage domains. If any subarray is not utilized for the layer, the power for the voltage domain is not supplied. As a result, we can maximize the energy-efficiency of the accelerator.

\section{Conclusion}
% In this paper, we present a dense-matrix multiplication accelerator based computing for the sparse neural networks, Sparse-on-Dense. Different from the sparse neural network accelerators adopting high-complexity circuits for index matching, Sparse-on-Dense consists of low-complexity hardware by performing dense matrix multiplication after decompressing the sparse format data. The experiment showed that the dense matrix multiplication of sparse neural networks can be more efficient in area and energy for the real world workloads.
In this paper, we presented a Sparse-on-Dense hardware architecture for efficient sparse neural network computation. Conventional sparse neural network accelerators use a complex index-matching logic and large-sized FIFOs/buffers to maintain high multiplier array utilization. However, such an approach leads to significant overheads. In contrast, we proposed to use the simple systolic array structure for computing, and the compressed sparse formats to reduce memory accesses. Our design shows lower multiplier array utilization than the sparsity-aware hardware baselines, but achieves much higher performance on the real benchmark datasets. Experimental results show that the proposed design improves the throughput/area and the effective energy-efficiency by up to 11.9$\times$ and 3.2$\times$ over the baseline designs, respectively.

\bibliographystyle{IEEEtran}
\bibliography{refer}

@article{han2015deep,
  title={Deep compression: Compressing deep neural networks with pruning, trained quantization and huffman coding},
  author={Han, Song and Mao, Huizi and Dally, William J},
  journal={arXiv preprint arXiv:1510.00149},
  year={2015}
}

@article{cheng2017model,
  title={A survey of model compression and acceleration for deep neural networks},
  author={Cheng, Yu and Wang, Duo and Zhou, Pan and Zhang, Tao},
  journal={arXiv preprint arXiv:1710.09282},
  year={2017}
}

@inproceedings{isca_tpu,
  title={In-datacenter performance analysis of a tensor processing unit},
  author={Jouppi, Norman P and Young, Cliff and Patil, Nishant and Patterson, David and Agrawal, Gaurav and Bajwa, Raminder and Bates, Sarah and Bhatia, Suresh and Boden, Nan and Borchers, Al and others},
  booktitle={Proceedings of the 44th annual international symposium on computer architecture},
  pages={1--12},
  year={2017}
}

@inproceedings{han2017ese,
  title={Ese: Efficient speech recognition engine with sparse lstm on fpga},
  author={Han, Song and Kang, Junlong and Mao, Huizi and Hu, Yiming and Li, Xin and Li, Yubin and Xie, Dongliang and Luo, Hong and Yao, Song and Wang, Yu and others},
  booktitle={Proceedings of the 2017 ACM/SIGDA International Symposium on Field-Programmable Gate Arrays},
  pages={75--84},
  year={2017}
}

@article{parashar2017scnn,
  title={Scnn: An accelerator for compressed-sparse convolutional neural networks},
  author={Parashar, Angshuman and Rhu, Minsoo and Mukkara, Anurag and Puglielli, Antonio and Venkatesan, Rangharajan and Khailany, Brucek and Emer, Joel and Keckler, Stephen W and Dally, William J},
  journal={ACM SIGARCH Computer Architecture News},
  volume={45},
  number={2},
  pages={27--40},
  year={2017},
  publisher={ACM New York, NY, USA}
}

@article{zhang2020snap,
  title={SNAP: An Efficient Sparse Neural Acceleration Processor for Unstructured Sparse Deep Neural Network Inference},
  author={Zhang, Jie-Fang and Lee, Ching-En and Liu, Chester and Shao, Yakun Sophia and Keckler, Stephen W and Zhang, Zhengya},
  journal={IEEE Journal of Solid-State Circuits},
  volume={56},
  number={2},
  pages={636--647},
  year={2020},
  publisher={IEEE}
}

@article{sanh2020movement,
  title={Movement pruning: Adaptive sparsity by fine-tuning},
  author={Sanh, Victor and Wolf, Thomas and Rush, Alexander M},
  journal={arXiv preprint arXiv:2005.07683},
  year={2020}
}

@article{han2015learning,
  title={Learning both weights and connections for efficient neural networks},
  author={Han, Song and Pool, Jeff and Tran, John and Dally, William J},
  journal={arXiv preprint arXiv:1506.02626},
  year={2015}
}

@article{brown2020language_gpt3,
  title={Language models are few-shot learners},
  author={Brown, Tom B and Mann, Benjamin and Ryder, Nick and Subbiah, Melanie and Kaplan, Jared and Dhariwal, Prafulla and Neelakantan, Arvind and Shyam, Pranav and Sastry, Girish and Askell, Amanda and others},
  journal={arXiv preprint arXiv:2005.14165},
  year={2020}
}

@article{rajpurkar2016squad,
  title={Squad: 100,000+ questions for machine comprehension of text},
  author={Rajpurkar, Pranav and Zhang, Jian and Lopyrev, Konstantin and Liang, Percy},
  journal={arXiv preprint arXiv:1606.05250},
  year={2016}
}

@unpublished{wang2018glue,
 title={GLUE: A Multi-Task Benchmark and Analysis Platform for
         Natural Language Understanding},
 author={Wang, Alex and Singh, Amanpreet and Michael, Julian and Hill,
         Felix and Levy, Omer and Bowman, Samuel R.},
 note={arXiv preprint 1804.07461},
 year={2018}
}

@article{krizhevsky2012alexnet,
  title={Imagenet classification with deep convolutional neural networks},
  author={Krizhevsky, Alex and Sutskever, Ilya and Hinton, Geoffrey E},
  journal={Advances in neural information processing systems},
  volume={25},
  year={2012}
}

@article{simonyan2014vgg,
  title={Very deep convolutional networks for large-scale image recognition},
  author={Simonyan, Karen and Zisserman, Andrew},
  journal={arXiv preprint arXiv:1409.1556},
  year={2014}
}

@article{naumov2019dlrm,
  title={Deep learning recommendation model for personalization and recommendation systems},
  author={Naumov, Maxim and Mudigere, Dheevatsa and Shi, Hao-Jun Michael and Huang, Jianyu and Sundaraman, Narayanan and Park, Jongsoo and Wang, Xiaodong and Gupta, Udit and Wu, Carole-Jean and Azzolini, Alisson G and others},
  journal={arXiv preprint arXiv:1906.00091},
  year={2019}
}

@article{devlin2018bert,
  title={Bert: Pre-training of deep bidirectional transformers for language understanding},
  author={Devlin, Jacob and Chang, Ming-Wei and Lee, Kenton and Toutanova, Kristina},
  journal={arXiv preprint arXiv:1810.04805},
  year={2018}
}

@inproceedings{qin2020sigma,
  title={Sigma: A sparse and irregular gemm accelerator with flexible interconnects for dnn training},
  author={Qin, Eric and Samajdar, Ananda and Kwon, Hyoukjun and Nadella, Vineet and Srinivasan, Sudarshan and Das, Dipankar and Kaul, Bharat and Krishna, Tushar},
  booktitle={2020 IEEE International Symposium on High Performance Computer Architecture (HPCA)},
  pages={58--70},
  year={2020},
  organization={IEEE}
}

@article{liu2020sta,
  title={Systolic tensor array: An efficient structured-sparse GEMM accelerator for mobile CNN inference},
  author={Liu, Zhi-Gang and Whatmough, Paul N and Mattina, Matthew},
  journal={IEEE Computer Architecture Letters},
  volume={19},
  number={1},
  pages={34--37},
  year={2020},
  publisher={IEEE}
}

@inproceedings{liu2022s2ta,
  title={S2ta: Exploiting structured sparsity for energy-efficient mobile cnn acceleration},
  author={Liu, Zhi-Gang and Whatmough, Paul N and Zhu, Yuhao and Mattina, Matthew},
  booktitle={2022 IEEE International Symposium on High-Performance Computer Architecture (HPCA)},
  pages={573--586},
  year={2022},
  organization={IEEE}
}

%\newpage

%\section{Biography Section}
%If you have an EPS/PDF photo (graphicx package needed), extra braces are
% needed around the contents of the optional argument to biography to prevent
% the LaTeX parser from getting confused when it sees the complicated
% $\backslash${\tt{includegraphics}} command within an optional argument. (You can create
% your own custom macro containing the $\backslash${\tt{includegraphics}} command to make things
% simpler here.)
 
%\vspace{11pt}

%\bf{If you include a photo:}\vspace{-33pt}
%\begin{IEEEbiography}[{\includegraphics[width=1in,height=1.25in,clip,keepaspectratio]{fig1}}]{Michael Shell}
%Use $\backslash${\tt{begin\{IEEEbiography\}}} and then for the 1st argument use $\backslash${\tt{includegraphics}} to declare and link the author photo.
%Use the author name as the 3rd argument followed by the biography text.
%\end{IEEEbiography}

%\vspace{11pt}

%\bf{If you will not include a photo:}\vspace{-33pt}
%\begin{IEEEbiographynophoto}{John Doe}
%Use $\backslash${\tt{begin\{IEEEbiographynophoto\}}} and the author name as the argument followed by the biography text.
%\end{IEEEbiographynophoto}

%\vfill

\begin{IEEEbiography}[{\includegraphics[width=1in,height=1.25in,clip,keepaspectratio]{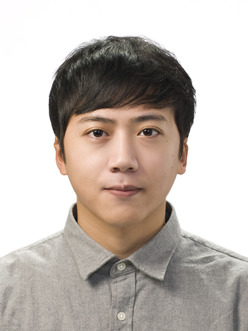}}]{Hyunsung Yoon} received the B.S. degree in electronic and electrical engineering from Chung-Ang University, Seoul, South Korea, in 2019. He received the Ph.D degree with the department of convergence IT engineering, Pohang University of Science and Technology, Pohang, South Korea in 2026. His current research interests include hardware architecture for deep neural networks, and sparse processing.
\end{IEEEbiography}

\begin{IEEEbiography}[{\includegraphics[width=1in,height=1.25in,clip,keepaspectratio]{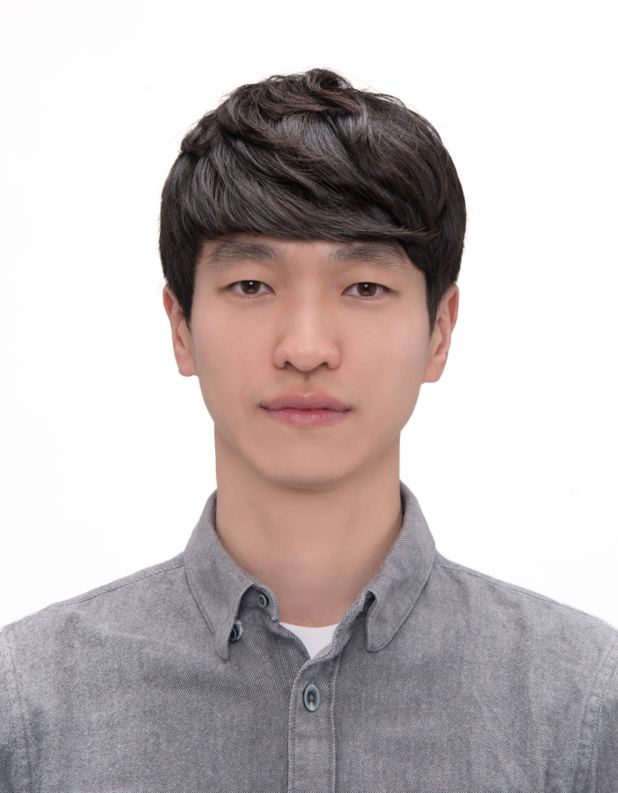}}]{Sungju Ryu}
(S'16-M'21) is currently an Associate Professor in the Department of System Semiconductor Engineering at Sogang University, Seoul, Republic of Korea. Before joining Sogang, he was an Assistant Professor in School of Electronic Engineering and Department of Next-Generation Semiconductor at Soongsil University from 2021 to 2023. At 2021, he was a Staff Researcher in the AI\&SW Research Center of Samsung Advanced Institute of Technology (SAIT), Suwon, Republic of Korea. At SAIT, he focused on computer architecture design. He received the B.S. degree in Electrical Engineering from Pusan National University, Busan, Republic of Korea, in 2015, and the Ph.D. degree in Creative IT Engineering from Pohang University of Science and Technology (POSTECH), Pohang, Republic of Korea, in 2021. His current research interests include energy-efficient hardware accelerators for AI applications, high-performance computing, and in-/near-memory computing.
\end{IEEEbiography}

\begin{IEEEbiography}[{\includegraphics[width=1in,height=1.25in,clip,keepaspectratio]{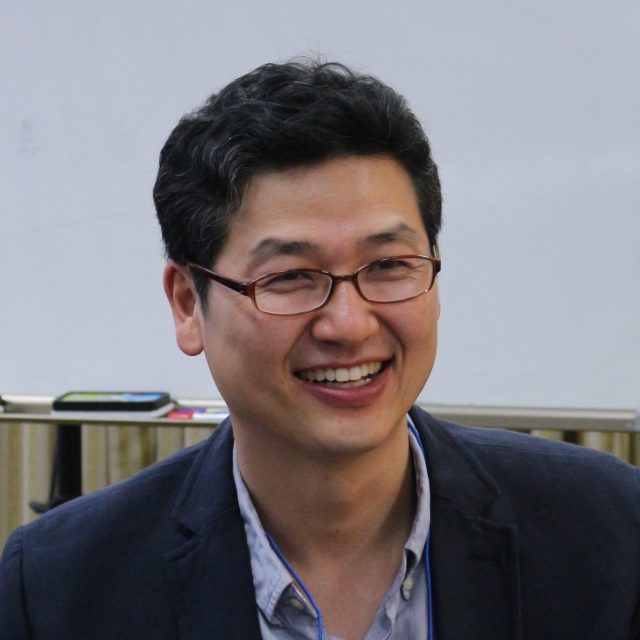}}]{Jae-Joon Kim}
(M'04) is currently a professor at Seoul National University, Seoul, South Korea. He received the B.S. and M.S. degrees in electronics engineering from Seoul National University, Seoul, South Korea, in 1994 and 1998, respectively, and the Ph.D. degree from the School of Electrical and Computer Engineering, Purdue University, West Lafayette, IN, USA, in 2004. From 2004 to 2013, he was a Research Staff Member with the IBM Thomas J. Watson Research Center, Yorktown Heights, NY, USA. He was a Professor at the Pohang University of Science and Technology, Pohang, South Korea from 2013 to 2021. His current research interests include the design of deep learning hardware accelerators, neuromorphic processors, hardware security circuits, and circuits for exploratory devices.
\end{IEEEbiography}

\vfill
\end{document}